\documentclass{acm_proc_article-sp}

\usepackage{listings}
\usepackage{xcolor}
\usepackage{subfig}
\usepackage[export]{adjustbox}
\usepackage{algorithmicx}
\usepackage{algpseudocode}
\usepackage{algorithm}

\usepackage{multirow}

\definecolor{codegreen}{rgb}{0,0.6,0}
\definecolor{codegray}{rgb}{0.5,0.5,0.5}
\definecolor{codepurple}{rgb}{0.58,0,0.82}
\definecolor{backcolour}{rgb}{0.95,0.95,0.92}

\colorlet{punct}{red!60!black}
\definecolor{background}{HTML}{EEEEEE}
\definecolor{delim}{RGB}{20,105,176}
\colorlet{numb}{magenta!60!black}

\lstdefinelanguage{json}{
    basicstyle=\normalfont\ttfamily,
    numbers=left,
    numberstyle=\scriptsize,
    stepnumber=1,
    numbersep=8pt,
    showstringspaces=false,
    breaklines=true,
    frame=lines,
    backgroundcolor=\color{background},
    literate=
     *{0}{{{\color{numb}0}}}{1}
      {1}{{{\color{numb}1}}}{1}
      {2}{{{\color{numb}2}}}{1}
      {3}{{{\color{numb}3}}}{1}
      {4}{{{\color{numb}4}}}{1}
      {5}{{{\color{numb}5}}}{1}
      {6}{{{\color{numb}6}}}{1}
      {7}{{{\color{numb}7}}}{1}
      {8}{{{\color{numb}8}}}{1}
      {9}{{{\color{numb}9}}}{1}
      {:}{{{\color{punct}{:}}}}{1}
      {,}{{{\color{punct}{,}}}}{1}
      {\{}{{{\color{delim}{\{}}}}{1}
      {\}}{{{\color{delim}{\}}}}}{1}
      {[}{{{\color{delim}{[}}}}{1}
      {]}{{{\color{delim}{]}}}}{1},
}

\begin{document}

\title{SDC-based Resource Constrained Scheduling for Quantum Control Architectures}
%
%
%
%
%

\numberofauthors{1} 
%
\author{
%
%
\alignauthor
R\u{a}zvan Nane\\
       \affaddr{Big Data Accelerate B.V.}\\
       \affaddr{Delft, The Netherlands}\\
       \email{razvan.nane@bigdataccelerate.com}
}

\maketitle
\begin{abstract}

Instruction scheduling is a key transformation in backend compilers that take an untimed description of an algorithm and assigns time slots to the algorithm's instructions so that they can be executed as efficiently as possible while taking into account the target processor limitations, such as the amount of computational units available. For example, for a superconducting quantum processor these restrictions include the amount of analogue instruments available to play the waveforms to drive the qubit rotations or on-chip connectivity between qubits. Current small-scale quantum processors contain only a few qubits; therefore, it is feasible to drive qubits individually albeit not scalable. Consequently, for NISQ and beyond NISQ devices, it is expected that classical instrument sharing to be designed in the future quantum control architectures where several qubits are connected to an instrument and multiplexing is used to activate only the qubits performing the same quantum operation at a time. Existing quantum scheduling algorithms either rely on ILP formulations, which do not scale well, or use heuristic based algorithms such as list scheduling which are not versatile enough to deal with quantum requirements such as scheduling with exact relative timing constraints between instructions, situation that might occur when decomposing complex instructions into native ones and requiring to keep a fixed timing between the primitive ones to guarantee correctness. In this paper, we propose a novel resource constrained scheduling algorithm that is based on the SDC formulation, which is the state-of-the-art algorithm used in the reconfigurable computing. We evaluate it against a list scheduler and describe the benefits of the proposed approach. We find that the SDC-based scheduling is not only able to find better schedules, with an improvement of 10\% on average, but it is also more versatile being able to model the complex relative timing constraints required for quantum circuit resource constrained scheduling. 

\end{abstract}


\section{Introduction}

Quantum technology promises to boost the computing capabilities available today by orders of magnitude, which will revolutionize key application domains and give birth to new ones that will drive the human evolution for decades to come. However, because of the novelty of the computing approach that completely differs from any existing classical computing technology, we require a holistic research and development agenda in which everything from the lowest physical level, the qubit, to the highest application level needs to be (re)invented. A key component in the quantum full stack is the (backend) compiler. The main objective of a compiler is to efficiently translate a quantum high-level algorithm into an optimal quantum circuit that can be executed correctly on a quantum chip. For example, quantum circuits, which are composed of a series of quantum operations called gates, require scheduling to ensure all the gates of the quantum algorithm and their dependencies are satisfied while making sure that the resource limitations of the target quantum chip are taken into consideration. Moreover, in the context of quantum compilation, where short decoherence times are an additional burden \cite{nielsen2002quantum}, the availability of a performant scheduling algorithm can be the difference between a correctly executing quantum algorithm and a completely useless circuit. 

At the same time, the number of qubits available in current quantum processors is low (at the moment of publication the largest device is IBM Eagle \cite{ibm:eagle} with 127 qubits), which implies that qubits can be driven individually and independently of other qubits because in these early quantum computing chips the quantum control electronics are not shared. Nevertheless, this simplistic approach is not realistic for scalable quantum computing systems starting with several hundreds to thousands of qubits, such as the IBM System Two quantum architecture \cite{ibm:systemtwo}, that will require multiplexing qubit control wires. Consequently, developing resource constrained scheduling algorithms is key to the success of these future architectures. However, existing approaches for current quantum computing deal with the scheduling problem in a trivial manner by mostly ignoring architectural limitations and resort to scheduling the quantum algorithm in an as soon as possible (ASAP) style, where only the program dependencies are taken into consideration. 

One of the few compiler frameworks for quantum computation that addresses the limitations mentioned above is OpenQL \cite{paper:openql}, which includes a backend list scheduling compiler pass \cite{openqldoc} for the Surface-17 superconducting quantum processor \cite{paper:s17}. However, the problem with using list scheduling in a quantum compiler backed is that it is not versatile enough to model quantum gate decomposition \cite{openqlissues:decomposition} and satisfy the relative timings finer quantum operations should obey with respect to one another after decomposition, e.g., performing a flux operation while at the same time parking other qubits, or ensuring fixed timing that might be required for feedback control or error detection and correction. The alternative to use a fully specified integer linear programming (ILP) formulation would solve the above \textit{versatility} problem, albeit not being scalable. Consequently, a scheduling algorithm that provides a balanced trade-off between versatility and scalability was proposed. This scheduling algorithm is based on a system of difference constraints (SDC) \cite{paper:sdc} formulation stemming from ILP and is the state-of-the-art in high-level synthesis compilers \cite{paper:surveyhls} used in classical reconfigurable computing. However, SDC scheduling with resource constraints in a quantum context is not optimal due to the way quantum resources are shared, i.e., one quantum instrument can perform multiple quantum gates of the same type at once similar to classical vector processing units.   

In this paper, we propose a novel quantum resource constraint scheduling algorithm based on SDC (QSDC) to generate efficient quantum schedules when quantum resources are shared. Concretely, the novelties of this paper are:
\vspace{-10pt}
\begin{itemize}
    \item We develop a novel resource constraint scheduling algorithm based on the SDC formulation and integrate it into the OpenQL compiler framework. 
    \item We provide a comprehensive analysis of the advantages of our proposed algorithm when compared with the current list scheduling algorithm available in OpenQL.  
\end{itemize}

\vspace{-10pt}
The paper is organized as follows. First, section \ref{section:background} presents the necessary background, including SDC preliminaries and related works. Then, in section \ref{section:qsdc} we present the QSDC algorithm. Section \ref{section:results} describes the experimental results. Finally, section \ref{section:conclusion} summarizes the paper and highlights future work.

\section{Background}
\label{section:background}

In this section, we introduce first the underlying concepts of quantum computation, then we present the instruction scheduling problem based on the system of difference constraints formulation, and finally, we review the scheduling state-of-the-art for quantum compilers.

\subsection{Quantum Computing and Resources}
\label{relres:qcs}
Quantum computing requires a radically novel approach to the developing of processors and compilers, the hardware and software building blocks of any computing system. The main reason for the requirement of new processor design techniques and novel compiler algorithms is due to the switch to implementing quantum mechanics operations rather than the classical approach of performing Boolean logic arithmetic. While processing units enabling Boolean logic operations can be implemented fully in the digital domain by mature EDA techniques using large-scale integrated circuits, quantum mechanics requires the integration of analogue instruments that drive the basic quantum computational unit, the qubit, by generating waveforms to instruct which quantum gate has to be performed on a qubit. Table \ref{table:classicalquantumcomparison} summarizes the differences between the classical and quantum basic computational concepts and processor micro-architecture functional units. 

\begin{table}[htbp]
\caption{Comparison of Classical vs. Quantum Computing}
\resizebox{\columnwidth}{!}{%
\begin{tabular}{|l|c|c|c|}
\hline
Compute Concept  &  Classical & Quantum \\ \hline\hline
Basic Compute Element   & bit & qubit \\ \hline
Logic Type & arith/boolean & quantum mechanics  \\ \hline
Logic Operations & +,-,*,|,\& & X, H, CNOT \\ \hline
Micro Architecture & only digital & digital \& analogue \\ \hline
\textit{enabled by an} & CU+ALU & CU+AWG \\ \hline
\end{tabular}
}
\label{table:classicalquantumcomparison}
\end{table}

\begin{figure}
\centering
\includegraphics[width=85mm]{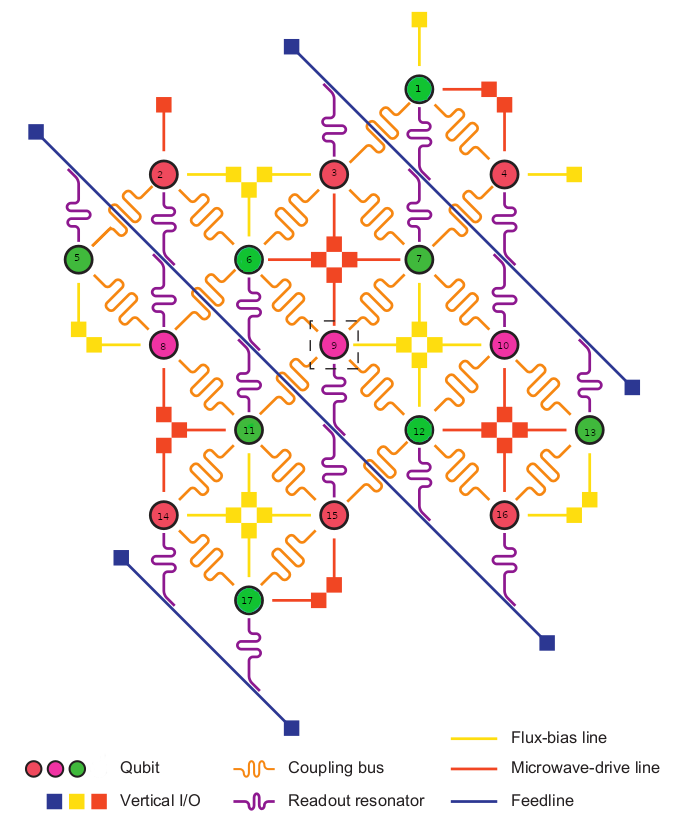}
\caption{Qutech Surface-17 (S-17) Quantum Processor. \\Figure edited from \cite{paper:s17}. }
\label{fig:s17}
\end{figure}

The major difference stems from the requirement to implement quantum mechanics that requires driving analogue devices, e.g., an Arbitrary Waveform Generator (AWG) to play different waveforms corresponding to particular quantum gates (e.g., an X gate, as opposed to an arithmetic operation performed by an Arithmetic-Logic Unig (ALU) in a classical processor) controlled by general-purpose Control Units (CUs) that keep track of which quantum gates have to be performed at a given time step according to the quantum algorithm. Consequently, due to the digital-analog domain crossing required in the design of a quantum processor micro-architectures, the sharing of AWGs is key to the success of developing scalable quantum computing systems. For example, Figure \ref{fig:s17} shows the Surface-17 \textit{"schematic of the targeted realization of Surface-17 in a planar cQED architecture with vertical I/O. Every transmon (represented by a circle) has dedicated flux control line, microwave-drive line, and readout resonator. Dedicated bus resonators mediate interactions between nearest-neighbor data and ancilla qubits. Readout resonators are simultaneously interrogated using frequency-division multiplexing in diagonally-running feedlines} \cite{paper:s17}. In the current S-17 configuration, qubits colored the same are connected to the same microwave-drive line and controlled by the same AWG instrument. For example, qubits \textit{8, 9, and 10} are driven by a single AWG. 

In this work, we use this target processor with the instrument connections depicted in Figure \ref{fig:s17}. However, the work can be easily retartgeted by modifying the instrument section in the OpenQL's platform configuration file, as shown in Listing \ref{lst:platconfig}. \textit{qwgs} section describe the connections for the single-qubit rotation gates (instructions of 'mw' type) that are controlled by AWGs.  Each qwg controls a private set of qubits, enumerated in the \textit{connection\_map}.  A qwg can control multiple qubits at the same time, but only when they perform the same gate and started at the same time. There are 'count' qwgs. For each qwg it is described which set of qubits it controls as configured for the S17 quantum device. Additionally, single-qubit measurements (instructions of 'readout' type) are controlled by measurement units.  Each one controls a private set of qubits. A measurement unit can control multiple qubits at the same time, but only when they start at the same time. There are 'count' measurement units and for each measurement unit it is described which set of qubits it controls. Sections concerning the available instructions and chip topology information related to the connectivity of the device is left out for space reason. 

\begin{lstlisting}[language=json,firstnumber=1,caption={Excerpt of the Platform Configuration File used by the OpenQL Compiler. Instrument Counts and Connectivity Correspond to the S-17 Shown in Figure \ref{fig:s17}.},label={lst:platconfig}]
"resources":
{
 "qubits": 
 {
  "count": 17
 },
 "qwgs":
 {
  "count": 3,
  "connection_map":
  {
   "0" : [2, 3, 4, 14, 15, 16],
   "1" : [8, 9, 10],            
   "2" : [1, 5, 6, 7, 11, 12, 13, 17]
  }
 },
 "meas_units" :
 {
  "count": 3,
  "connection_map":
  {
   "0" : [14,17],
   "1" : [2, 5, 6, 8, 9, 11, 12, 15, 16],
   "2" : [1, 3, 4, 7, 10, 13]
  }
 }, ...%topology and instructions%..
}
\end{lstlisting}

\begin{figure}
\centering
\includegraphics[width=85mm]{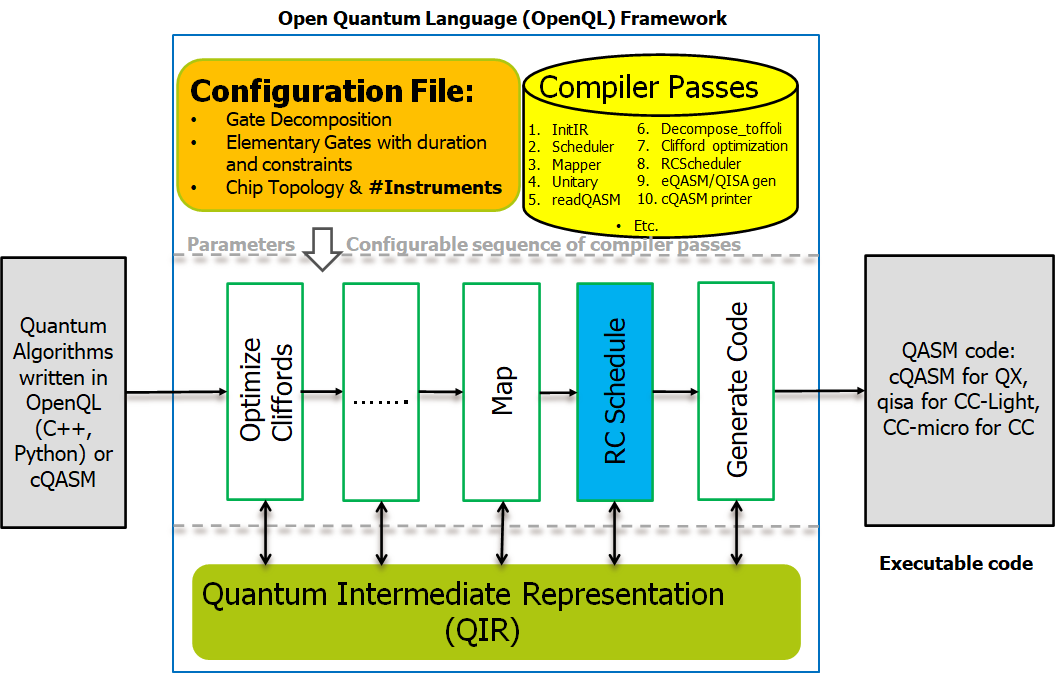}
\caption{OpenQL Quantum Compiler Framework Overview.}
\label{fig:openql}
\end{figure}

OpenQL \cite{paper:openql} is quantum compiler framework developed by Qutech, depicted graphically in Figure \ref{fig:openql}. It is a modular and retargetable framework as it allows to easily add new compiler passes as well as generate code for different quantum devices and technologies by simply describing the chip architecture in a new platform configuration file \cite{openqldoc}, illustrated above. OpenQL currently supports ASAP and ALAP scheduling, and resource constrained list scheduling. Furthermore, quantum programs can be written either in C++ or using the pyhon API. It can generate both simulatable cQASM code \cite{cqasm1:langauge} for QX simulator \cite{paper:qx} and quantum micro-code (CC-micro) for Qutech's Central Controller \cite{Moreira2019QuTechCC}.

\subsection{SDC Preliminaries}

Instruction scheduling is a central problem in compilers. Informally, the instruction scheduling problem can be formulated as finding the best (i.e., usually the one that runs the fastest) sequence of instructions that minimize the execution time of an algorithm given different constraints, such as the available resources existent in Instruction scheduling is a central problem in compilers. Informally, the instruction scheduling problem can be formulated as finding the best (i.e., usually the one that runs the fastest) sequence of instructions that minimize the execution time of an algorithm given different constraints, such as the available resources existent in the target processor. Formally, the scheduling problem can be defined as an assignment of execution slots to each instruction (i.e., the time when the instruction is active) so that all program and platform dependencies are taken into account. For example, using integer linear programming (ILP) the following constraint has to be imposed on the \textit{scheduling variables} defined for each instruction so that a valid schedule is produced:   

\vspace{-15pt}
\begin{align}
\sum_{j=1}^{m} x_{i}^{j} = 1
\end{align}
\vspace{-15pt}

where it is assumed an \textit{m} clock-cycle schedule for each of the \textit{i} instructions in the program.
According to \cite{paper:ilp}, equation (1) is called an \textit{appearance constraint} and is needed to ensure one instruction will only be executed in exactly one cycle.

\begin{figure}
\centering
\includegraphics[width=85mm]{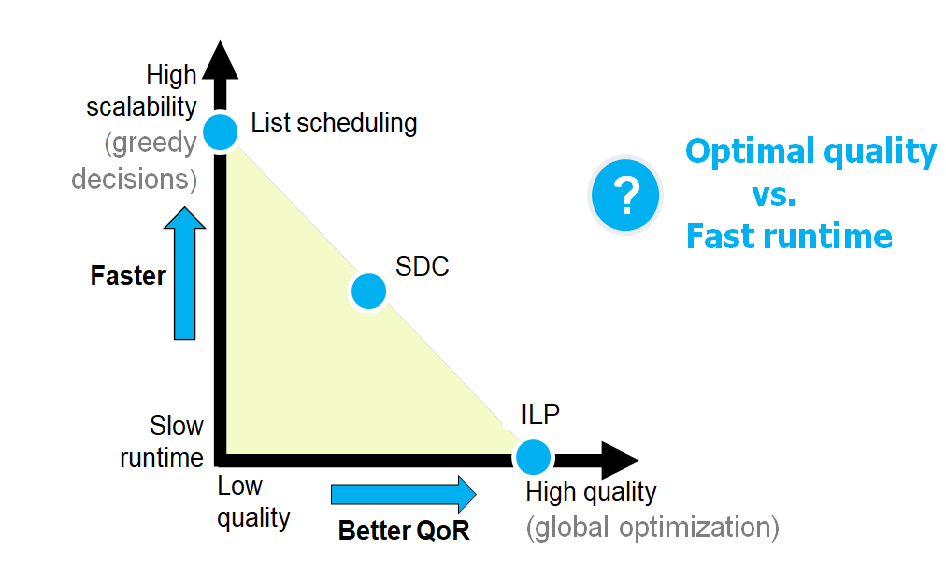}
\caption{Tradeoff between Scalability and Quality}
\label{fig:sdcvsilp}
\end{figure}

Several other constraints have to be formulated in a similar manner as in (1) to solve the scheduling problem. Although using this formal specification using a mathematical approach is optimal, giving the best Quality-of-Results (QoR), due to the difficulty of solving it for large problems (i.e., the scheduling problem under resource constraints is known to be NP-hard), alternative scheduling algorithms have been proposed that are based on heuristics. One well-known algorithm is list scheduling \cite{paper:listscheduling}, which uses a ready list of instructions and sorts them in increasing order of some predefined priority to select the next node to be scheduled. By using this ready list of instructions, the algorithm reduces the search space that in turn increases the scalability of the algorithm with the risk of obtaining a local minima solution; therefore, degrading the QoR of the obtained schedule. Figure \ref{fig:sdcvsilp} highlights this trade-off between QoR and scalability.

As a comprimise between the two orthogonal features, i.e., runtime vs. QoR, a more versatile scheduling heuristic has been proposed in \cite{paper:sdc} based on a system of integer difference constraints (SDC). This heuristic is rooted in ILP, however, due to the intelligent way of encoding the scheduling variables and instruction constraints, which are defined as a linear system of inequalities where the underlying matrix is \textit{totally unimodular}, the scheduling formulation can be solved using a linear programming relaxation that generates optimal integer solutions in polynomial time. Therefore, SDC can find better schedules than list scheduling while being faster than the fully specified ILP problem. Furthermore, contrary to list scheduling, in the SDC formulation we can easily specify \textit{relative timing constraints}, which are very important in quantum computing because it is often necessary to guarantee an exact latency distance between two operations that would avoid for example the situation where the qubits will decohere. Using the terminology in \cite{paper:sdc}, equations (2) and (3) highlight the inequalities needed to specify an exact timing constraint between two operations \textit{a} and \textit{b}: 

\vspace{-15pt}
\begin{align}
sv_{beg}(a) - sv_{beg}(b) \leq -l_{ab}
\end{align}
\vspace{-15pt}

\vspace{-15pt}
\begin{align}
sv_{beg}(b) - sv_{beg}(a) \leq l_{ab}
\end{align}
\vspace{-15pt}

, where $sv_{beg}$ is the first cycle scheduling variable associated with an instruction and $l_{ab}$ is the number of clock cycles between \textit{a} and \textit{b}. It is worth noting that several other constraints can be specified using the SDC formalism, all of which can be found in \cite{paper:sdc}.

\subsection{Related Works}      

Several quantum compilation frameworks have been developed over the last decade and in this part we will highlight some of them focusing on the available scheduling algorithm and target platform supported as a differentiating factor from this work. 
One of the first compilers developed was the ScaffCC \cite{scaffcc:compiler} compiler for the Scaffold programming language. Developed initially by the Princeton University, in collaboration with IBM T.J. Watson and University of Santa Barbara, the compiler was build using the LLVM compiler infrastructure \cite{llvm:compiler} and offered several advanced compiler transformations, such as the RKQC, the reversible logic circuitry toolkit for quantum computation, and the Longest-Path-First-Schedule (LPFS) scheduling compiler pass. Furthermore, another important characteristic of ScaffCC is that it is able to generate OpenQASM v2.0 \cite{openqasm2:language} and cQASM v1.0 \cite{cqasm1:langauge} quantum assembly languages. However, ScaffCC can be considered only a \textit{front-end} compiler because it does not support a target quantum processor, rather it defers this compilation process to a \textit{backend-compiler}, available for example in IBM's QisKit runtime \cite{}, that knows the resource limitation of that particular quantum device. Consequently, the LPFS scheduler in ScaffCC is just a variant of an As-Soon-As-Possible (ASAP) scheduler that does not consider any resource constraints. 

Qiskit \cite{qiskit:compiler} is another compilation framework developed by IBM. The software development kit is written in python for fast prototyping and uses a list scheduler for the different backends it supports. However, due to the limited amount of qubits available in the early quantum devices, i.e., up to 127 qubits available in the IBM Eagle that was based on the System One architecture, there was no multiplexing of the control wires of each qubits. Consequently, there was no need for advanced scheduling algorithms that optimize the circuit under resource sharing incurred by control wire stemming from multiplexing qubit control wires as required for developing scalable quantum control architectures starting with IBM System Two quantum architecture \cite{ibm:systemtwo}. Therefore, developing resource constrained scheduling algorithm is key to the success of these future architectures.  

Other compilers, such as Qcor \cite{qcor:compiler} and t|ket \cite{tket:compiler}, are suffering the same drawbacks as the compilers described above, namely they are focusing on front-end compilation tasks and defer backend target compilation to quantum device providers and their backend runtime software, e.g., via IBM Quantum Experience. The main limitation of this approach is, as previously mentioned, that due to current small size of existing quantum devices the scheduling algorithms used were mostly based on basic ASAP style of scheduling. Contrary to this state-of-the-art in quantum circuit scheduling, we focus on the resource constraint scheduling problem for future quantum processors, such as those based on IBM's System Two architecture, that will include sharing of quantum control electronics. We integrate our proposed QSDC scheduling algorithm into OpenQL compiler framework and target the Surface-17 chip that has a scalable architecture by sharing its control electronics as described in section \ref{relres:qcs}.

\section{QSDC Scheduling Algorithm}
\label{section:qsdc}

In this section, we will describe the QSDC scheduling algorithm at the hand of a simple example shown in Algorithm \ref{src:examplecircuit}. The quantum circuit is written using OpenQL's python API and is composed of four gates \textit{X,Y,X, and Z} that operate on three qubits \textit{2,3, and 4}. Recall that according to the S-17 instrument connections the qubits involved are driven by the same \textit{AWG} with id 0 (see Figure \ref{fig:s17} and Listing \ref{lst:platconfig}).

\begin{algorithm}
\caption{Running Example Quantum Circuit}
\label{src:examplecircuit}
...\\
k.gate(“x”, [2])\\
k.gate(“y”, [3])\\
k.gate(“x”, [4])\\
k.gate(“z”, [2])\\
...
\end{algorithm}

\begin{figure*}[t]
\centering
\subfloat[SDC ASAP Scheduling Formulation.]{
\includegraphics[width=.32\textwidth,valign=t]{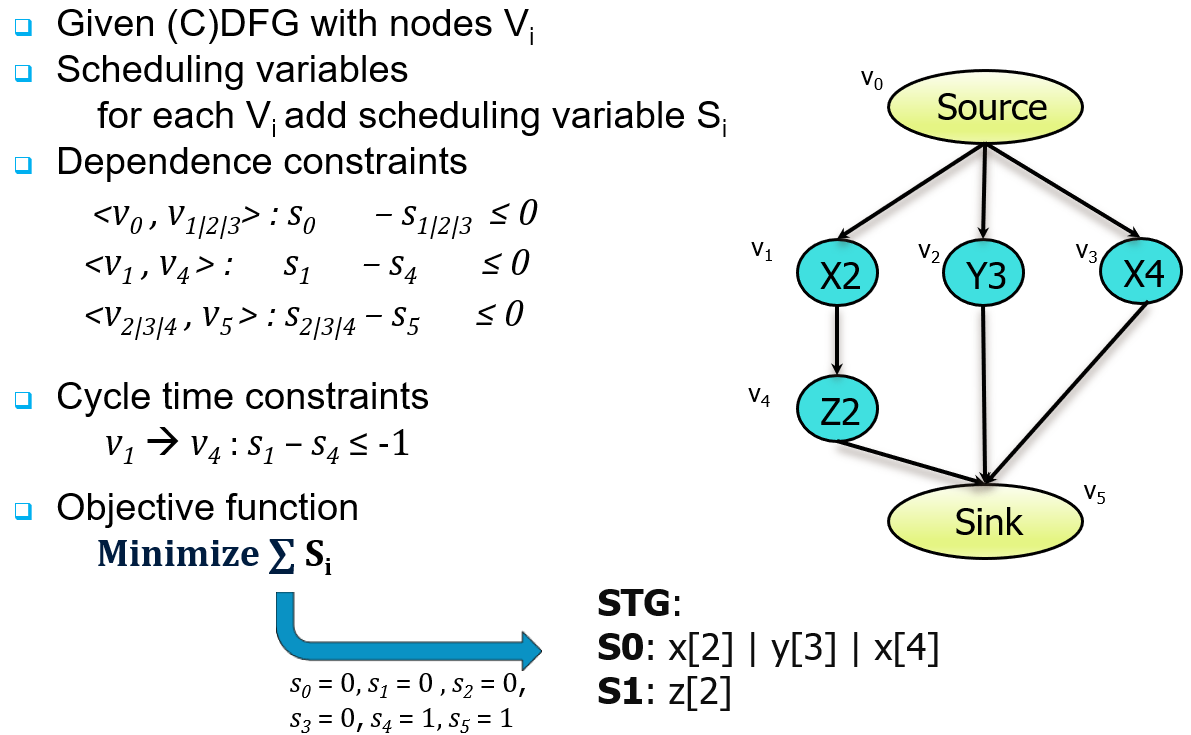}
\label{fig:linordera}
}
\subfloat[Default Linear Order for Adding\\ Resource Constraints as in \cite{paper:sdc}. \\Suboptimal schedule due to RC with \\latency $\leq$ -1 added between Y3 and X4.]{
\includegraphics[width=.32\textwidth,valign=t]{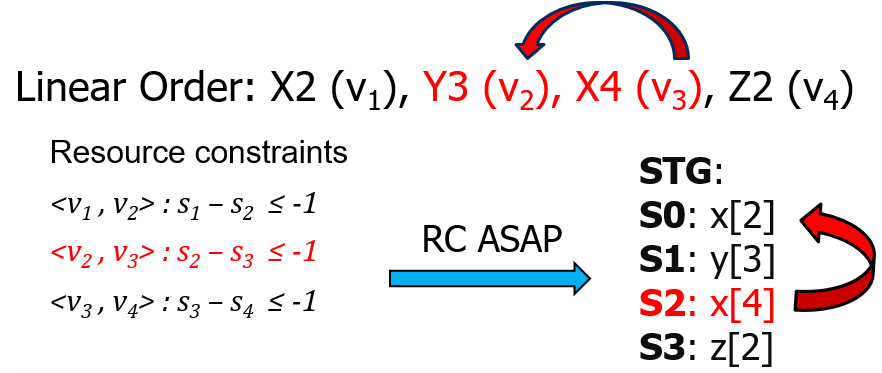}
\label{fig:linorderb}
}
\subfloat[Optimal QSDC Linear Order for Adding Resource Constraints. Y3 and X4 have to be interchanged to piggyback on X2 if possible (latency $\leq$ 0). In this example, X2 and X4 can run in parallel.]{
\includegraphics[width=.32\textwidth,valign=t]{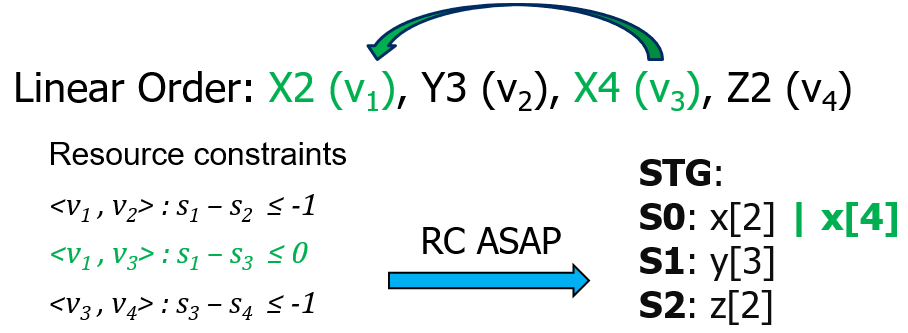}
\label{fig:linorderc}
}
\caption{Example CDFG, SDC Formulation, and Resosurce Constraints Linear Orders.}
\label{fig:linorder}
\end{figure*}

To compile any (quantum) program, a compiler will first create a control and data dependency graph(CDFG) out of the program instructions that fulfill the dependencies of the program code, e.g., the abstract operations X and Z should happen exactly in the order they appear in the code. Please note that we call these operations \textit{abstract} because at this point the compiler does not know that the quantum gates X and Z operating on qubit 2 are commutative. The (C)DFG for our example is shown in Figure \ref{fig:linordera}. Alongside we show the SDC formulation when the chosen scheduling option is As-Soon-As-Possible(ASAP) without considering any resource constraints. When we schedule in this way, we see that a state transition graph(STG) is created in which all the independent operations are scheduled in parallel in the first state. That is, the first cycle 0 is assigned to the \textit{scheduling variables} (s1, s2, and s3) belonging to the v1 to v4 graph nodes, while cycle 1 is assigned to s4 of the Z2's v4 node.

Before we continue with the formulation for the resource constrained scheduling problem, we recall that the scheduling problem with resource constraints is known to be an NP-hard problem. Therefore, to solve large problems, we rely on heuristics such as list scheduling or SDC, which also cannot model exactly the resource constraints. Consequently, the heuristic in SDC is to use a sorting algorithm to create a linear order of the CDFG nodes and then to use this order to add constraints based on the resources types and counts. In the best mode of operation known for SDC, as explained in [6], this includes sorting the nodes in ascending order using an As-Late-As-Possible (ALAP) primary key with the ASAP key as a tiebreaker when a couple of nodes have the same ALAP cycle. However, this is not optimal for quantum computing as illustrated in Figure \ref{fig:linorderb}. Using the X2, Y3, X4, Z2 linear order created by the aforementioned sorting heuristic and considering that an AWG can perform multiple operations in parallel only if they are the same type, a resource constraint between Y3 and X4 has to be imposed such that X4 starts at least 1 cycle after Y3 finishes because they both require the same AWG-0. Similarly, two other resource constraints have to be added between <v1 and v2> and <v3 and v4>, leading to a sub-optimal schedule in which all 4 nodes execute individually on the AWG requiring 4 cycles to run the whole quantum circuit.

However, if we consider that an AWG can control multiple qubits at the same time when they perform the same gate started at the same time, we observe that the X2 and X4 quantum gates could have been scheduled in parallel because \textit{stacking} is possible in a quantum context. This observation led us to introduce a new heuristic that we call QSDC, which can be combined with the default resource constraint linear order sorting heuristic to generate a more optimized STG that takes into consideration the quantum resource features. QSDC enables stacking of quantum operations on the same quantum resource whenever possible, as shown in Figure \ref{fig:linorderc}. To track the free stacking space an instrument has, we introduce the concept of a \textit{running instruction} that tracks if and how many instructions are assigned to which instruments and what is the maximum allowed stacking. Whenever we schedule an instruction we refer to the last running instruction of its type and \textit{anchor} to it by adding a resource dependency of \textit{$\leq$ 0} from its scheduling variable to the running instruction's variable, instead of the previously added \textit{$\leq -1$} dependency to the last instruction of a different type.

\begin{algorithm}
\caption{Main Resource Constrained QSDC}
\label{alg:qsdc1}
\begin{flushleft}
\textbf{Function}: $addResourceConstraints$ \\
\textbf{Input}: $CDFGNodes[]$, $PlatformResources[]$, $QSDC$ \\
\mbox{\Comment{Goal: Assign rc\_cycles to the CDFGNodes}}\\
\mbox{\Comment{Prerequisites: \{asap|alap\}\_cycles are valid}}\\
\textbf{Output}: $StateTransitionGraph[]$ \\
\mbox{\Comment{STG where rc\_cycles are assigned to CDFGNodes}}
\end{flushleft}
\begin{algorithmic}[1]                   
    \For{$resourceType : PlatformResources$} 
    \For{$resourceInstance : resourceType$} 
    \State $qubitConnections[] = getConnectedQubitsOf(resourceInstance)$
    \State $rcinstr.clear()$
    \For{$qubit : qubitConnections$} 
    \For{$gateOP : CDFGNodes$} 
    \If{$isQubitNeeded(gateOP, qubit)$}
    \State $rcinstr.push\_back(gateOP)$
    \State $break$
    \EndIf
    \EndFor
    \EndFor
    \EndFor
    \If {$(rcinstr.size() > 1)$}
    \State $addResourceConstraintsInstrument($\\$rcinstr,getInstrStacking(resourceInstance, QSDC))$
    \EndIf
    \EndFor
\end{algorithmic}
\end{algorithm}

The full QSDC RC scheduling algorithm is given in Algorithms \ref{alg:qsdc1} and \ref{alg:qsdc2}. In the first algorithm, the overall main loop is shown to iterate through the resource types (e.g., AWG or readout units) of the platform and for each type it selects each instance (e.g., 0, 1, or 2 for an AWG type), shown in lines 1-2. Then, the CDFG nodes using a qubit that is connected to the selected instrument and that performs a quantum gate supported by that instrument are added to a new list (lines 3-13). If any instructions are found, then the method to add resource constraints for that instruments is called (lines 14-17). This method is depicted in the second algorithm. Here, for each new instruction to be scheduled (line 3), two main cases are considered. First, if the instruction type is already \textit{running} (lines 4-6), then a $\leq 0$ inequality is added between the running instruction and this instruction (lines 7-11), unless the maximum stacking was reached, in which case a $\leq -1$ inequality is added and the counters are reset to 1 (lines 12-16). Second, if the instruction is not running, then the last instruction of other type is found (line 25) and a $\leq -1$ inequality is added to the scheduling formulation (line 26-29). In the case no running instruction is there (line 19), we simply initialize the running instruction list with this instruction (line 21) and set the last instruction see to this (line 30).

\begin{algorithm}[h]
\caption{Instrument Resource Constrained QSDC}
\label{alg:qsdc2}
\begin{flushleft}
\textbf{Function}: $addResourceConstraintsInstrument$ \\
\textbf{Input}: $instructions[]$, $maxStacking$, $QSDC$ \\
\mbox{\Comment{All instructions needing this instrument}}\\
\mbox{\Comment{Goal: set SDC constraints between instructions}}\\
\textbf{Output}: $QSDC\_RCextended$ \\
\mbox{\Comment{RC constraints are added to QSDC resulting in \_RCextended  }}
\end{flushleft}
\begin{algorithmic}[1]                    
    \State $schedInstr=\{firstSchedVar,lastSchedVar,stackCount\}$
    \State $Map<instrOP, schedInstr> runInstructions$
    \For{$instruction : instructions$} 
    \State $runInstr = runInstructions.find(instruction)$
    \If{$(runInstr != runInstructions.end())$}
\\\mbox{\Comment{instruction type is already running  }}
    \If{$(runInstr.stackCount < maxStacking)$}
\\\mbox{\Comment{instruction can be stacked, i.e., \textbf{<= 0}}}
    \State $rcinstr.push\_back(gateOP)$
    \State $addRCBetween(runInstr, instruction, 0)$
    \State $runInstr.stackCount++$
    \Else 
\\\mbox{\Comment{max stacking reached, i.e., \textbf{<= -1}}}
    \State $addRCBetween(runInstr, instruction, -1)$
    \State $runInstr.stackCount = 1$
    \EndIf
    \Else
\\\mbox{\Comment{instruction not running, and ...}}
    \If{(runInstructions.size() == 0)}
\\\mbox{\Comment{ ... and no other instr occupying the resource}}
     \State runInstructions \textbf{<-} instruction
    \Else
\\\mbox{\Comment{ ... but other instr occupying the resource,i.e., \textbf{<= -1}}}
    \State instr =\\ runInstructions.find(lastInstructionSeen);
    \State $addRCBetween(instr, instruction, -1)$
    \EndIf
    \State runInstructions[instruction].stackCount = 1
    \EndIf
    \State $lastInstructionSeen = instruction$
    \EndFor
\end{algorithmic}
\end{algorithm}

We integrated these algorithms in the OpenQL compiler as a new scheduling pass. To solve the linear problem relaxation we used the \textit{lpsolve} software package for solving linear, integer and mixed programs \cite{src:lpsolve}. 
\section{Experimental Results}
\label{section:results}

In this section, we evaluate the QSDC algorithm by compiling ten circuits from \cite{paper:qmap} for the superconducting processor Surface-17 that shares several control electronics among the 17 qubits for different types of quantum operations, limitations described in section \ref{relres:qcs} and highlighted in Figure \ref{fig:s17}. This sharing is translated into the resource constraints described in the platform configuration file used by the OpenQL compiler (see Listing \ref{lst:platconfig}). We implemented the QSDC compiler into the release version 0.8.1 of OpenQL and compare it against the default existing resource constrained list scheduling algorithm. Furthermore, because the list scheduling pass was recently updated in OpenQL version 0.10.5, the latest version at the time of writing, we compare against this as well. Please note that several bug fixes were done from version 0.8.1 to 0.10.5, which increased the overall latency of the compiled benchmarks. We compiled and run the benchmarks in an Ubuntu 20.04.5 LTS installation running on an eight-core eight-thread Intel(R) Core(TM) i7-9700 processor @ 3.00 GHz with 16 GB of memory.

\subsection{Benchmarks}
The ten benchmarks randomly selected from \cite{paper:qmap} are described in Table \ref{table:benchmarks}, which lists the initial number of qubits, gates, and CNOTs as specified in the original, not scheduled, circuit. To compile and schedule these circuits so that they can correctly execute on the S-17 quantum processor, the following compiler passes are selected, executed in the listed order: \textit{Clifford gate optimizer}, \textit{Qmap}, \textit{Clifford gate optimizer}, and \textit{RCsched}, which was configured both as a list scheduling and qsdc for the experiments. For the Qmap pass the \textit{minextendrc} option was set, while the platform configuration file did not include the \textit{gate\_decomposition} section. It is important to realize that each of these passes modify the original specification to optimize it as well as to solve connectivity limitations (i.e., by Qmap), which introduce additional gates to move the qubits into adjacent positions before the actual \textit{CNOTs} gates can be executed. For example, the Qmap compiler pass increases the number of quantum operation and therefore the latency of the circuit. The QSDC as well as the resource constraint scheduler are invoked after the qmap has generated a topologically correct circuit. 

\begin{table}[htbp]
\caption{Selected Quantum Circuits}
\resizebox{\columnwidth}{!}{%
\begin{tabular}{|l|c|c|c|c|}
\hline
Benchmark  &  Qubits & Gates & CNOTs & Latency \\ \hline
4gt12\_v1\_89  & 6 & 228 & 100 & 448 \\ \hline
4gt4\_v0\_72  & 6 & 258 & 113 & 478 \\ \hline
alu\_bdd\_288 & 7 & 84 & 38 & 169 \\ \hline
alu\_v0\_27 & 5 & 36 & 17 & 72 \\ \hline
clip\_206 & 14 & 33827 & 14772 & 61786 \\ \hline
mini\_alu\_167 & 5 & 288 & 126 & 564 \\ \hline
mini\_alu\_305 & 10 & 173 & 77 & 242 \\ \hline
sqrt8\_260 & 12 & 3009 & 1314 & 5740 \\ \hline
sym10\_262 & 12 & 64283 & 28084 & 122564 \\ \hline
z4\_268 & 11 & 3073 & 1343 & 5688 \\ \hline
\end{tabular}
}
\label{table:benchmarks}
\end{table}

\newpage
\subsection{QSDC Evaluation}
The compiler version used to obtain the results are 0.8.1 and 0.10.5 form the OpenQL github repository \cite{src:openqlgithub}. The results are shown in Table \ref{table:results}, where we can observe an increase in both number of qubits and gates used. The Latency column list the circuit execution time in cycles considering the real gate duration that are specified in the platform configuration file (20 ns per cycle). Finally, the two speed-up columns describe the speedup obtain of the QSDC algorithm against the default list scheduler when we compare to both the OpenQL versions tested. 

\begin{table}[t]
\caption{Experimental Evaluation Results}
\resizebox{\columnwidth}{!}{%
\begin{tabular}{|c|c|c|c|c|c|c|c|}
\hline
\multicolumn{1}{|c|}{Benchmark} & \multicolumn{1}{c|}{Version} & \multicolumn{ 1}{c|}{Qubits} & Gates & CNOTs & Latency & \multicolumn{2}{c|}{Speedup} \\ \hline\hline
\multirow{3}{*}{4gt12\_v1\_89} & 0.10.5 & 9 & 311 & 183 & 1361 & \multicolumn{1}{l|}{} & 1 \\ \cline{2-8}
\multicolumn{ 1}{|c|}{} & 0.8.1 & 9 & 310 & 182 & 873 & 1 & \multicolumn{1}{l|}{} \\ \cline{ 2- 8}
\multicolumn{ 1}{|c|}{} & Qsdc-0.8.1 & 10 & 308 & 180 & 809 & \textbf{1.08} & \textbf{1.68} \\ \hline \hline
\multirow{3}{*}{4gt4\_v0\_72} & 0.10.5 & 9 & 327 & 182 & 1202 & \multicolumn{1}{l|}{} & 1 \\ \cline{ 2- 8}
\multicolumn{ 1}{|c|}{} & 0.8.1 & 10 & 340 & 195 & 953 & 1 & \multicolumn{1}{l|}{} \\ \cline{ 2- 8}
\multicolumn{ 1}{|c|}{} & Qsdc-0.8.1 & 10 & 340 & 195 & 875 & \textbf{1.09} & \textbf{1.37} \\ \hline \hline
\multirow{3}{*}{alu\_bdd\_288} & 0.10.5 & 9 & 115 & 69 & 485 & \multicolumn{1}{l|}{} & 1 \\ \cline{ 2- 8}
\multicolumn{ 1}{|c|}{} & 0.8.1 & 10 & 114 & 68 & 341 & 1 & \multicolumn{1}{l|}{} \\ \cline{ 2- 8}
\multicolumn{ 1}{|c|}{} & Qsdc-0.8.1 & 8 & 116 & 70 & 329 & \textbf{1.04} & \textbf{1.47} \\ \hline \hline
\multirow{3}{*}{alu\_v0\_27} & 0.10.5 & 7 & 51 & 32 & 210 & \multicolumn{1}{l|}{} & 1 \\ \cline{ 2- 8}
\multicolumn{ 1}{|c|}{} & 0.8.1 & 9 & 50 & 31 & 147 & 1 & \multicolumn{1}{l|}{} \\ \cline{ 2- 8}
\multicolumn{ 1}{|c|}{} & Qsdc-0.8.1 & 5 & 45 & 26 & 131 & \textbf{1.12} & \textbf{1.60} \\ \hline \hline
\multirow{3}{*}{clip\_206} & 0.10.5 & 17 & 45265 & 26210 & 169513 & \multicolumn{1}{l|}{} & 1 \\ \cline{ 2- 8}
\multicolumn{ 1}{|c|}{} & 0.8.1 & 16 & 45521 & 26466 & 114880 & 1 & \multicolumn{1}{l|}{} \\ \cline{ 2- 8}
\multicolumn{ 1}{|c|}{} & Qsdc-0.8.1 & 16 & 45368 & 26313 & 110965 & \textbf{1.04} & \textbf{1.53} \\ \hline \hline
\multirow{3}{*}{mini\_alu\_167} & 0.10.5 & 9 & 368 & 206 & 1384 & \multicolumn{1}{l|}{} & 1 \\ \cline{ 2- 8}
\multicolumn{ 1}{|c|}{} & 0.8.1 & 9 & 371 & 209 & 1020 & 1 & \multicolumn{1}{l|}{} \\ \cline{ 2- 8}
\multicolumn{ 1}{|c|}{} & Qsdc-0.8.1 & 8 & 362 & 200 & 940 & \textbf{1.09} & \textbf{1.47} \\ \hline \hline
\multirow{3}{*}{mini\_alu\_305} & 0.10.5 & 13 & 234 & 138 & 790 & \multicolumn{1}{l|}{} & 1 \\ \cline{ 2- 8}
\multicolumn{ 1}{|c|}{} & 0.8.1 & 10 & 226 & 130 & 467 & 1 & \multicolumn{1}{l|}{} \\ \cline{ 2- 8}
\multicolumn{ 1}{|c|}{} & Qsdc-0.8.1 & 10 & 243 & 147 & 486 & \textbf{0.96} & \textbf{1.63} \\ \hline \hline
\multirow{3}{*}{sqrt8\_260} & 0.10.5 & 13 & 3977 & 2282 & 15067 & \multicolumn{1}{l|}{} & 1 \\ \cline{ 2- 8}
\multicolumn{ 1}{|c|}{} & 0.8.1 & 13 & 3987 & 2292 & 10298 & 1 & \multicolumn{1}{l|}{} \\ \cline{ 2- 8}
\multicolumn{ 1}{|c|}{} & Qsdc-0.8.1 & 13 & 4050 & 2355 & 10191 & \textbf{1.01} & \textbf{1.48} \\ \hline \hline
\multirow{3}{*}{sym10\_262} & 0.10.5 & \multicolumn{4}{c|}{SEG FAULT} & \multicolumn{1}{l|}{} & 1 \\ \cline{ 2- 8}
\multicolumn{ 1}{|c|}{} & 0.8.1 & 13 & 85750 & 49551 & 221390 & 1 & \multicolumn{1}{l|}{} \\ \cline{ 2- 8}
\multicolumn{ 1}{|c|}{} & Qsdc-0.8.1 & 13 & 85702 & 49503 & 214126 & \textbf{1.03} & \#NA \\ \hline \hline
\multirow{3}{*}{z4\_268} & 0.10.5 & 13 & 4041 & 2311 & 15279 & \multicolumn{1}{l|}{} & 1 \\ \cline{ 2- 8}
\multicolumn{ 1}{|c|}{} & 0.8.1 & 13 & 4099 & 2369 & 10651 & 1 & \multicolumn{1}{l|}{} \\ \cline{ 2- 8}
\multicolumn{ 1}{|c|}{} & Qsdc-0.8.1 & 13 & 4038 & 2308 & 10017 & \textbf{1.06} & \textbf{1.53} \\ \hline 
\end{tabular}
}
\label{table:results}
\end{table}

Please note that the goal of the evaluation is focused on the sdc formulation in comparison with list scheduling. We did not intend to compare the efficiency of different implementations of list scheduling against other, since this is tightly connected to the platform and the resources. For example, in Scaffcc there is a LPFS scheduler that deal with resources;
however, this is not modeling a chip, rather a fictional one where all qubits are driven individually. 
\section{Conclusions and Future Work}
\label{section:conclusion}

In this paper, we proposed a novel resource constrained scheduling algorithm that is based on the SDC formulation, which is the state-of-the-art algorithm used in the reconfigurable computing. Concretely, we have proposed a new selection heuristic to select the operation nodes in a quantum efficient way to account for the quantum context where quantum gates can be stacked on the same instrument when they are the same type. Furthermore, we have validated our work by integrating the QSDC algorithm into the OpenQL compiler framework and evaluated it using ten quantum algorithms. The experimental results showed the benefits of QSDC against the default OpenQL list scheduling, where an average of 10\% speedup was observed. In future work, we will analyze and demonstrate the benefits of controlling relative timing constraints between quantum operations. However, before that can be showed, support for integrating flexible conditions between quantum gates have to be added at the data-dependecy graph in OpenQL.

\bibliographystyle{unsrt}
\bibliography{sigproc}  
\end{document}